# Time dilation in relativistic two-particle interactions

B.T. Shields[1], M.C. Morris[1], M.R. Ware[1],
Q. Su[1], E.V. Stefanovich[2] and R. Grobe[1]

(1) Intense Laser Physics Theory Unit and Department of Physics
Illinois State University, Normal, IL 61790-4560 USA

(2) 2255 Showers Drive, Unit 153, Mountain View, CA 94040 USA

We study the orbits of two interacting particles described by a fully relativistic classical mechanical Hamiltonian.  We use two sets of initial conditions. In the first set (dynamics 1) the system's center of mass is at rest. In the second set (dynamics 2) the center of mass evolves with velocity V.  If dynamics 1 is observed from a reference frame moving with velocity –V, the principle of relativity requires that all observables must be identical to those of dynamics 2 seen from the lab frame.  Our numerical simulations demonstrate that kinematic Lorentz space-time transformations fail to transform particle observables between the two frames. This is explained as a result of the inevitable interaction-dependence of the boost generator in the instant form of relativistic dynamics. In spite of general inaccuracies of Lorentz formulas, the orbital periods are correctly predicted by the Einstein's time dilation factor for all interaction strengths.

# 1. Introduction

The fundamental principle of relativity requires that the laws of physics should be invariant under changes of inertial reference frames. This property is formulated in terms of symmetry of the theory under the Poincaré group of inertial transformations [1-3]. For simplicity, in this work we will limit ourselves to one spatial dimension only. Then inertial observers are related to each other by space, time and velocity translations associated with the three generators P, H, and K, respectively. In classical mechanics, they have to satisfy the three Poisson brackets (Lie algebra)

$$\{P,H\} = 0 \qquad (1.1a)$$
$$\{H,K\} = P \qquad (1.1b)$$
$$\{P,K\} = H/c^2 \qquad (1.1c)$$

If we would like to predict how an observable A(X,P) (as a function of the phase-space variables X and P) is measured from a different reference frame, we have to solve the equation

$$\partial A(s)/\partial s = \{G, A(s)\} \qquad (1.2)$$

with the initial condition A(s=0)=A(X,P), where G is either P, K or H and {…, …} denotes the Poisson bracket. Each generator is associated with its corresponding group parameter s, which can be either the displacement d, the rapidity $c\theta$ (where $\theta=\tanh^{-1}(V/c)$ is a function of the velocity V), or the time t of the new reference frame relative to the lab frame.

In the instant form of relativistic dynamics, both the Hamiltonian H and the velocity boost operator K must depend on the interaction in order to satisfy the Poincaré relations [2-5]. Therefore one can expect that boost transformations of particle trajectories are interaction-dependent and system-specific [6, 7]. This conclusion is in obvious disagreement with the traditionally assumed interaction-independent and universal form of the position-time Lorentz transformation formulas (x, t) → (x′, t′)

$$x' = x \cosh(\theta) - c\, t \sinh(\theta) \qquad (1.3a)$$
$$t' = x \sinh(\theta)/c - t \cosh(\theta) \qquad (1.3b)$$

To study potential deviations from predictions of Eqs. (1.3), several works [8-11] have investigated the decay law of an unstable particle acting like a moving clock. Using relativistic quantum mechanics, they suggested that the true decay law for a fast particle is not provided by the Einstein's time dilation factor. Unfortunately, the predicted deviations are several orders of magnitude smaller than the experimental error of the most accurate experiments to date [12-13].

In this work we consider another type of physical system, where deviations from Eqs. (1.3) can be, in principle, observed. We give a concrete example for a system of two mutually attracting classical particles. As these two particles oscillate with respect to each other, there are distinct periodic moments in time when their trajectories cross each other. The occurrences of these unambiguous events serve as a clock, and we can test how Einstein's time dilation formula would describe the observations by a moving observer. In agreement with theoretical predictions [6], our numerical simulations demonstrate that space-time Lorentz transformations Eqs. (1.3) do not hold for particle trajectories. Moreover, strong interaction potentials allow for particle velocities higher than the speed of light. Nevertheless, Einstein's time dilation formula remains valid for the orbital periods.

## 2. Non-interacting particles

Let us consider a system of two classical spinless particles. We assume that they have the same mass $m_1=m_2\equiv m$, which will make the expressions below a little bit more transparent. The phase space associated with positions $x_i$ and momenta $p_i$ for $i=1,2$ is four-dimensional and we define the usual Poisson brackets as $\{A, B\} = \Sigma_i \partial_{x_i}A\partial_{p_i}B - \partial_{x_i}B\partial_{p_i}A$. If the particles are non-interacting, then Poincaré relations (1.1) are easily satisfied by choosing

$$P = p_1+p_2 \qquad (2.1a)$$
$$H_0 = h_1+h_2 \qquad (2.1b)$$
$$K_0 = k_1+k_2 \qquad (2.1c)$$

where $h_i \equiv (m_i^2 c^4 + c^2 p_i^2)^{1/2}$, $k_i = -x_i h_i/c^2$ and the subscript "0" indicates the absence of interactions. It is convenient to introduce the total mass observable M (which has vanishing Poisson brackets with all three generators P, $H_0$, $K_0$) and the center of mass position R

$$M = [H_0^2 - c^2 P^2]^{1/2}/c^2 \qquad (2.2)$$
$$R = (x_1 h_1 + x_2 h_2)/(h_1+h_2) \qquad (2.3)$$

Then Eqs. (2.1b) and (2.1c) can be re-written as

$$H_0 = [M^2 c^4 + c^2 P^2]^{1/2} \qquad (2.4)$$
$$K_0 = - R\, H_0/c^2 \qquad (2.5)$$

## 3. Transformations of observables between different frames

In this section we will provide two examples for using the Poincaré generators (2.1) constructed above to connect observables measured in different inertial frames, corresponding to a *passive* coordinate transformation. These examples are combined to derive the usual Lorentz formulas (1.3) for non-interacting particle systems. The corresponding equivalent *active* transformations, where the coordinates are shifted by d, cθ or t can be obtained by reversing the sign of the parameter. For example, if in Eq. (1.2) we choose A=$x_i$ or $p_i$ with G=$H_0$ and s=-t, then we obtain the familiar Hamilton equations of motion for the time-dependence of particle positions and momenta

$$\partial x_i(t)/\partial t = -\{H_0, x_i(t)\} = \partial H_0/\partial p_i = c^2 p_i(t)/h_i(t) = v_i(t) \quad (3.1a)$$
$$\partial p_i(t)/\partial t = -\{H_0, p_i(t)\} = -\partial H_0/\partial x_i = 0 \quad (3.1b)$$

where velocities are defined as $v_i = \partial x_i(t)/\partial t$. These equations have simple solutions

$$p_i(t) = p_i(0) \quad (3.1c)$$
$$x_i(t) = x_i(0) + v_i(0)t \quad (3.1d)$$

corresponding to freely propagating particles, as expected.

For a second example we consider the passive boost transformations of the total momentum and energy. In this case we use Eq. (1.2) with A= P or $H_0$ and G=$K_0$ with s=cθ. The resulting two coupled differential equations are

$$\partial P(\theta)/\partial(c\theta) = -H_0(\theta)/c^2 \quad (3.2a)$$
$$\partial H_0(\theta)/\partial(c\theta) = -P(\theta) \quad (3.2b)$$

Their solution leads to the well-known formulas

$$P(\theta) = P(0) \cosh(\theta) - H_0(0) \sinh(\theta)/c \quad (3.2c)$$
$$H_0(\theta) = H_0(0) \cosh(\theta) - c P(0) \sinh(\theta) \quad (3.2d)$$

Similarly one can obtain transformation laws for one-particle momenta and energies

$$p_i(\theta) = p_i(0) \cosh(\theta) - h_i(0) \sinh(\theta)/c \quad (3.3a)$$
$$h_i(\theta) = h_i(0) \cosh(\theta) - c p_i(0) \sinh(\theta) \quad (3.3b)$$

From these equations we obtain the usual relativistic velocity addition law

$$v_i(\theta) = c^2 p_i(\theta)/h_i(\theta) = [v_i(0) - c\,\tanh(\theta)]/[1 - v_i(0)\tanh(\theta)/c] \tag{3.4}$$

noting that $c\tanh(\theta)$ is the relative velocity of the moving frame. Let us now calculate boost transformations for positions. From Eq. (1.2) with $A=x_i$ and $G=K_0$, $s=c\theta$ we obtain equation

$$\partial x_i(\theta)/\partial(c\theta) = \{-x_1(\theta)h_1(\theta) - x_2(\theta)h_2(\theta), x_i(\theta)\}/c^2 = x_i(\theta)v_i(\theta)/c^2 \tag{3.5a}$$

which, together with the initial condition $x_i(\theta=0)=x_i(0)$, results in the familiar length contraction

$$x_i(\theta) = x_i(0)h_i(0)/h_i(\theta) \tag{3.5b}$$

The time evolution of the particle position in the moving frame is a function of time $t'$ measured by the moving clock

$$x_i(\theta,t') = x_i(\theta) + v_i(\theta)t' = x_i(0)h_i(0)/h_i(\theta) + c^2 p_i(\theta)/h_i(\theta)\,t' \tag{3.6}$$

Suppose that observer at rest sees the i-th particle at location $x_i(t)$ at (lab time) t. Let us define a specific time $t'$ (measured by the clock in the moving frame) by the requirement

$$t' \equiv t\cosh(\theta) - x_i(t)\sinh(\theta)/c \tag{3.7a}$$

Can we find the associated position where the particle is located from the point of view of the moving observer at this time? To do that, we replace in the right hand-side of Eq. (3.6) $x_i(0)$ by $x_i(t)-v_i(0)t$, and $t'$ by Eq. (3.7a), and use $c^2 p_i(0) = v_i(0)h_i(0)$ from Eq. (3.1a). The right hand side of Eq. (3.6) can then be expressed in terms of the original observable $x_i(0, t)$ as

$$x_i(\theta,t') = x_i(0,t)[h_i(0)/h_i(\theta) - cp_i(\theta)\sinh(\theta)/h_i(\theta)] - t[c^2 p_i(0)/h_i(\theta) - c^2 p_i(\theta)\cosh(\theta)/h_i(\theta)]$$
$$= x_i(0,t)\cosh(\theta) - ct\sinh(\theta) \tag{3.7b}$$

which we recognize as the Lorentz formula (1.3) for the coordinate. In other words, we have demonstrated that traditional Lorentz transformations for non-interacting particles can be derived directly from the three Poincaré relations by first applying a passive boost based on K to the initial position and momentum which is then followed by the (active) time evolution with the Hamiltonian $H(\theta)$ transformed to the new frame. We note that only this particular sequence of two actions reproduces the Lorentz formulas and that the derivation is only valid for non-interacting particles.

## 4. Quasi-relativistic approximation

Results from the preceding section encourage us to seek transformation formulas for interacting particles as well. In this paper we work exclusively in the instant form of relativistic dynamics where interaction enters in H and K, while the generator of space translations remains interaction-free. Before dealing with a rigorously relativistic system in the next section, here we would like to consider a quasi-relativistic approximation in which an expansion is made in powers of the small parameter $c^{-2}$ and all terms smaller than $c^{-2}$ are omitted. In this approximation the non-interacting generators (2.1b) - (2.1c) can be written as

$$H_0 = 2mc^2 + p_1^2/(2m) + p_2^2/(2m) + O[c^{-2}] \qquad (4.1)$$
$$K_0 = -m(x_1+x_2) - (p_1^2 x_1 + p_2^2 x_2)/(2m c^2) + O[c^{-4}] \qquad (4.2)$$

Note that even though $O[c^{-2}]$ terms in (4.1) are omitted, the above generators (together with P) satisfy Poincaré brackets (1.1) to the order $c^{-2}$. Within the same quasi-relativistic approximation interacting generators can be chosen as [14]

$$H_{qr}(x_1, x_2, p_1, p_2) = 2mc^2 + p_1^2/(2m) + p_2^2/(2m) + U(x_1-x_2) \qquad (4.3)$$
$$K_{qr}(x_1, x_2, p_1, p_2) = -m(x_1+x_2) - [(p_1^2 x_1 + p_2^2 x_2)/m + (x_1+x_2) U(x_1-x_2)]/(2 c^2) \qquad (4.4)$$

where an arbitrary function $U(x_1-x_2)$ has been introduced to serve the role of the potential energy of the interaction. Despite the interaction, the three Poincaré brackets Eq. (1.1) can be verified up to the order $c^{-2}$.

Let us now see how this interaction potential modifies the Lorentz transformation rules. Using the transformation equation (1.2) for the velocity translation we need to solve the four coupled partial differential equations for i=1,2

$$\partial x_i(\theta)/\partial(c\theta) = \{K_{qr}, x_i\} = -\partial K_{qr}(\theta)/\partial p_i = p_i x_i/m c^2 \qquad (4.5a)$$
$$\partial p_i(\theta)/\partial(c\theta) = \{K_{qr}, p_i\} = \partial K_{qr}(\theta)/\partial x_i$$
$$= -m - [p_i^2/m + U(x_1-x_2) + (x_1+x_2) \partial U(x_1-x_2)/\partial x_1]/(2 c^2) \qquad (4.5b)$$

For the specific initial condition $x_i(\theta=0)=0$, equations (4.5a) yield $x_i(\theta)=0$. The two momenta equations (4.5b) become decoupled and can be solved analytically too. In the lowest order in $c^{-2}$ and approximating the rapidity $\theta \approx V/c$, we obtain

$$p_i(\theta) \approx p_i(0) - mV - U(r=0)V/(2c^2) \tag{4.6}$$

The second term is the Galilei transformation and the third term is correction associated with the interaction. We expect this correction to have a significant effect if the interaction is strong, $|U(0)| > 2mc^2$.

Similar to the boost transformations of $x_i$ and $p_i$ described above, all Eqs. from (3.1) to (3.7) in Sec. 3 should be modified in the presence of interactions. This conclusion, however, does not refer to the boost transformations of the total momentum and energy in Eq. (3.2). These formulas are always valid, whether or not the particles interact, because they follow directly from the fundamental Poincaré brackets (1.1), which are not altered by the presence of interactions.

## 5. The relativistically invariant Bakamjian-Thomas interaction

The necessity to modify Lorentz transformations (3.7) in the presence of interactions is well-known. In 1963 Currie, Jordan and Sudarshan [6] proved a theorem stating that in relativistic Hamiltonian theories Lorentz formulas can be valid only in the non-interacting case. Our goal in the rest of this paper is to illustrate this remarkable theorem for a classical mechanical system of two interacting particles.

Constructing a relativistically invariant interacting model is a rather non-trivial task. In 1953, Bakamjian and Thomas [15] suggested an ingenious method for solving this problem. In our one-dimensional case their method amounts to finding a function $U(x_1, x_2, p_1, p_2)$ that has vanishing Poisson brackets with the total momentum and the center-of-mass position

$$\{R, U\} = \{P, U\} = 0 \tag{5.1}$$

This "interaction potential" U can be used to replace the mass observable $Mc^2$ in the Hamiltonian H and in the corresponding velocity boost generator K by $Mc^2 + U$. The new mass inserted in definitions Eqs. (2.4) and (2.5) gives

$$H = [(Mc^2 + U)^2 + c^2 P^2]^{1/2} \tag{5.2}$$
$$K = -R H/c^{-2} \tag{5.3}$$

Then it is not difficult to prove that the defined generators of time translations and boosts together with the non-modified generator of space translations P exactly satisfy the fundamental brackets Eqs. (1.1).

The practical way to construct a Bakamjian-Thomas interaction U that satisfies Eq. (5.1) is first to define the canonical pair of relative position and momentum variables r and p, which fulfill the relations {R,P}={r,p}=1 and {R,r}={R,p}={P,r}={P,p}=0. Here we choose the relative coordinates in the form [16]

$$r = (x_1-x_2) + (x_1-x_2) P^2 [(h_1+h_2+Mc^2)^{-1} - 4p^2(h_1+h_2)^{-1} M^{-2}c^{-2}]/M \qquad (5.4a)$$

$$p = (p_1-p_2)/2 - P(h_1-h_2)/(h_1+h_2+Mc^2)/2 \qquad (5.4b)$$

In the non-relativistic (Galilean) limit $c \to \infty$ these complicated expressions take the much more familiar forms $r \to (x_1-x_2)$, $p \to (p_1-p_2)/2$. Any reasonable function U(r,p) of these variables immediately satisfies conditions (5.1) and thus can be used in the Bakamjian-Thomas construction. In what follows we will consider only interactions U(r) that do not depend on the relative momentum p and vanish at large relative distances $U(r \to \infty)=0$.

It turns out that the mass function M depends only on the relative momentum, $M = 2(m^2c^4 + c^2p^2)^{1/2}/c^2$. This simple equality follows after a complicated algebra if we insert the expressions for $h_1$ and $h_2$ into Eq. (5.4), and solve this equation for the mass. Then the Bakamjian-Thomas Hamiltonian simplifies to

$$H = \left[\left(2(m^2c^4 + c^2p^2)^{1/2} + U(r)\right)^2 + c^2P^2\right]^{1/2} \qquad (5.5)$$

## 6. Numerical results in the lab frame

In order to study the time evolution and to go beyond the non-relativistic limit, we have to solve the system numerically. For this task and also to establish the complicated Poisson brackets above we used the Mathematica software package that allows for advanced symbolic manipulations as well as numerical solutions. We used the attractive Coulomb-like potential $U(r) = U_0/[r^2+a^2]^{1/2}$, where the singularity has been removed at the screening length a. From now on we use atomic units in which c=137.036 and m=1. Parameters of the interaction potential were chosen as a=1 and $U_0=-5c^2$. This means that condition $|U(0)| > 2mc^2$ is satisfied and we can expect to see a large effect of the interaction on the boost transformations. We solved the corresponding four coupled Hamilton equations of motion numerically for two sets of initial conditions both of which are characterized by $x_1(t=0)=x_2(t=0)=0$.

In the first case, we chose the initial momenta $p_1(t=0)=-400$ and $p_2(t=0)=400$ such that the center of mass remains at the origin, R=P=0. As our interaction potential is negative, the two particles are attracted to each other and periodically pass through each other. The corresponding orbits are shown in Fig. 1a. We note that the two orbits cross each other $x_1(t_n)=x_2(t_n)$ at characteristic times $t_n = 9.22 \times 10^{-2}$ n (for n=0,1,2, ...).

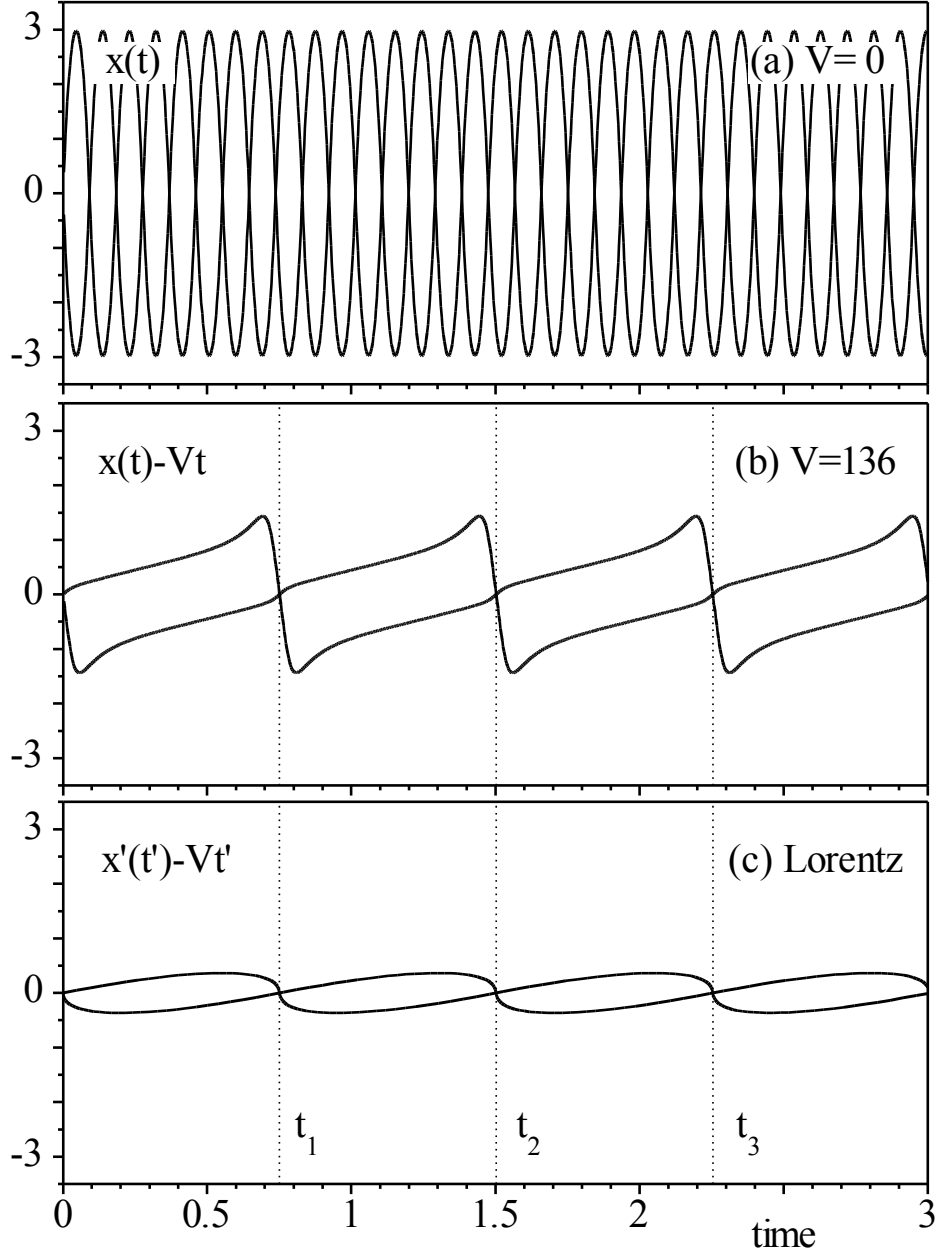

Figure 1: Two-particle trajectories $x_1(t)$ and $x_2(t)$ as a function of time
(a) for $p_1(t=0)= -400$ and $p_2(t=0)=400$ (corresponding to $v_1(t=0)=-129.64$ and $v_2(t=0)= 129.64$)
(b) for $p_1(t=0)= -83.82$ and $p_2(t=0)=1381.39$ (corresponding to $v_1(t=0)= 94.12$ and $v_2(t=0)= 140.85$), the drift term $Vt$ has been subtracted out.
(c) the Lorentz transformed trajectories based on the solutions displayed in (a). The drift term $Vt'$ has been subtracted out. [Parameters in all simulations: $x_1(t=0)=x_2(t=0)=0$, $m_1=m_2=1$, $c=137.036$, $U_0=-5c^2$, $a=1$]

The second set of initial momenta $p_1(\theta, t=0)$ and $p_2(\theta, t=0)$ was obtained by actively boosting the original momenta by the positive velocity, $V=136$ ($\theta=2.79$ and $\cosh(\theta)=8.15$). The numerical values for the new initial momenta were obtained by solving the boost equations

$$\partial x_i(\theta)/\partial(c\theta) = \{K(\theta), x_i(\theta)\} \tag{6.1a}$$
$$\partial p_i(\theta)/\partial(c\theta) = \{K(\theta), p_i(\theta)\} \tag{6.1b}$$

with the full interaction-dependent boost generator (5.3). This leads to initial positions $x_1(\theta, 0) = x_2(\theta, 0)=0$ and momenta $p_1(\theta, 0)= -83.82$ and $p_2(\theta, 0)=1381.4$. In this case the orbits also exhibit mutual oscillations, but since the center of mass is not at rest, there is an additional drift present. For better graphical clarity, in Fig. 1b we have subtracted this constant drift from the two orbits, $x_i(t,\theta) - Vt$. The asymmetric shape of the orbits might seem unusual. However, in contrast to the non-relativistic Hamiltonian (4.3), where the potential U depends exclusively on the inter-particle positions $x_1(t)-x_2(t)$, the relativistic potential energy is a very complicated function of $x_1$, $x_2$, $p_1$ and $p_2$. This momentum dependence makes it difficult to interpret the orbits within customary non-relativistic concepts.

The characteristic crossing times can be read off the graph as $t_n = 0.7514$ n. The fast moving clock seems to tick 8.15 times slower than the clock at rest. This effect would not be predicted by the non-relativistic Hamiltonian (4.3), for which the dynamics of the relative and center-of-mass coordinates are completely decoupled.

We should also comment on the unusual relationship between the canonical momenta and velocities. The velocities (defined as $v_i=\{x_i,H\}=\partial H/\partial p_i$) are non-monotonic functions of $p_i$ due to the relativistic interaction. The first set of momenta $p_i=\pm 400$ is associated with velocities $v= \pm 129.64$, while the second set corresponds to velocities $v_1= 94.12$ and $v_2=140.85$. As the velocity $v_2$ is larger than c, it is obvious that the usual velocity addition formula (3.4) is not valid in the interacting case.

## 7. Predictions by the Lorentz formulas

Let us next examine how the usual space-time Lorentz formulas (1.3) would predict the time evolution seen from a frame that is moving with a negative velocity, $V = -136$. If these formulas were applicable, the result of their application to the first set of orbits [$x_i(0)=0$, $p_i(0)=\pm 400$] would have to be identical to the trajectories obtained in our second simulation (Fig. 1b). We inserted the numerical solutions $x_i(t)$ shown Fig. 1a into Eqs. (1.3) and thus obtained the Lorentz transformed orbits $x_i'(t')$. These special-relativistic predictions are displayed in Fig. 1c. For better graphical clarity we have again removed the overall drift. In obvious violation of the principle of relativity, these trajectories are quite different from those shown in Fig. 1b. The most obvious difference is in the maximum amplitude (excursion) of the two sets of trajectories. For the original set of orbits (Fig. 1a) the oscillation amplitude is 2.974. In contrast, the correct amplitude for the moving system in Fig. 1b is contracted by a factor of 2.07 to the value of 1.435. The amplitude in Fig. 1c is shrunk 8.15 times with respect to Fig. 1a. This corresponds exactly to Einstein's length contraction factor of cosh(θ)=8.15.  This discrepancy shows that the usual length contraction formula does not work in the presence of interactions.

From these results it should be clear that Lorentz transformations (1.3) cannot be used to compute the evolution in a velocity boosted reference frame.  The correct approach would be to passively transform the initial conditions from $x_i(0,0)$ and $p_i(0,0)$ to $x_i(\theta, 0)$ and $p_i(\theta, 0)$ by solving the coupled equations (6.1) with θ=-2.79.  Due to the inherent equivalency of active transformations by V=136 and passive transformations by V=–136, this procedure leads exactly to the initial values used for Fig. 1b.  Then these four initial conditions need to be evolved in time under the boost-transformed Hamiltonian $H(\theta) = H\big(r(\theta), p(\theta), P(\theta)\big)$, thus resulting in the same time evolution as obtained in the lab frame for the second set of initial momenta (shown in Fig.1b) and restoring our confidence in the principle of relativity.

## 8. Validity of the time dilation formula for orbital periods

In the preceding section we have established that transforming particle orbits by the standard Lorentz formulas (1.3) leads to wrong results. However, there is one aspect of these transformations, which works surprisingly well.  According to the Lorentz formula, the transformed orbits exhibit particle crossings ($x_1'(t_n')=x_2'(t_n')$) at times $t_n'= t_n \cosh(\theta) = 0.7514$ n. This period of oscillations is exactly the same as in the correct result (compare graphs 1b and 1c). In other words, Einstein's time dilation formula

$$t_n(\theta)=t_n(0) \cosh(\theta) \qquad (8.1)$$

is still valid even if the two particles interact. This property is not an accident, but a rigorous result.

The proof is based on the universal (interaction-independent) relations (3.2).  From Eq. (5.5), the orbit of the moving system $r(p,P(\theta))$ in terms of the relative coordinates r and p and total energy $E(\theta)$ is defined as the solution to

$$2(m^2c^4 + c^2p^2)^{1/2}+U(r) = E(\theta)^2 - c^2P(\theta)^2 = E(0)^2 \qquad (8.2)$$

where the latter equality follows from the boost invariance of the mass-squared function $(Mc^2 + U)^2= E(\theta)^2 - c^2P(\theta)^2$.  The equation of motion for the relative momentum $dp/dt=\{p,H\}=-\partial H/\partial r$, can be integrated in time from t=0 to the first crossing time $t_1$, and correspondingly in momentum from p(0) to $p(t_1)=-p(0)$. For the system at rest this leads to

$$t_1(0)= - \int_{p(0)}^{-p(0)} dp \, (\partial U(r(p))/\partial r)^{-1} \qquad (8.3)$$

Using Eq. (5.5) we find for the moving system

$$t_1(\theta) = - \int_{p(0)}^{-p(0)} dp \ E(\theta)/\left[\left(2(m^2c^4 + c^2p^2)^{1/2} + U(r(p))\right) \partial U(r(p))/\partial r\right] \tag{8.4}$$

In the special case of $x_1 = x_2$ the relative momentum p is invariant under velocity boosts: $\{K,p\}=0$, therefore the initial condition p(0) (and thus the limits of integration in (8.4)) does not depend on $\theta$. If we further use the first equality in (8.2), expression (8.4) simplifies to

$$t_1(\theta) = -E(\theta)/[E(\theta)^2 - c^2 P(\theta)^2]^{1/2} \int_{p(0)}^{-p(0)} dp \ [\partial U(r(p))/\partial r]^{-1} \tag{8.5}$$

From the last equality in (8.2), the prefactor of the integral becomes $E(\theta)/E(0) = \cosh(\theta)$ leading to the final result (8.1).

## 9. Summary and brief discussion

We have modeled a classical mechanical clock by two particles bound to each other by an attractive potential. The time interval between particle crossings is a natural period of the clock. We found that usual space-time Lorentz formulas cannot describe the internal dynamics of this system in the moving frame. Nevertheless, the increased period of the moving clock is fully consistent with Einstein's time dilation formula. This is interesting as particle velocities can even exceed the speed of light. The key for understanding these unusual effects is the fact that the generator of velocity boosts K is interaction-dependent.

The two-particle model investigated in this work is open to some objections. One could argue that this model is unphysical because it involves instantaneous action-at-a-distance, which is known to violate causality. One could also argue that a more realistic analysis of Lorentz transformations should be performed within quantum field theory where interactions are transmitted by force-intermediating subluminal virtual particles and the concept of action at a distance is not required. However, these objections do not look undisputable to us.

First, the claims of causality violations by instantaneous interactions [17] are based on the validity of usual Lorentz formulas (1.3). However, as we have shown here, these formulas are no longer accurate in the presence of interactions. If the interaction dependence of boost transformations is taken into account, then it can be shown [7] that the cause and effect remain simultaneous in all inertial frames, and the causality is not necessarily violated.

Second, the speed of propagation of interactions in quantum field theory is still an unsettled issue [18, 19]. In fact, the entire QFT framework is oriented toward scattering problems and is not designed to answer questions about the time evolution and boost transformations of particle observables in the interaction zone. The visualization of interactions in terms of exchanges of virtual particles is non-trivial as these particles are not directly observable. In the dressed particle approach to quantum field theories [7, 20-23], physical particles interact with each other directly without virtual intermediaries. Nevertheless, the physical requirements of relativistic invariance and causality are satisfied. The possible violation of the Lorentz transformation formulas for interacting particles is obviously a very fundamental and non-trivial issue [24, 25] and deserves further discussions.

**Acknowledgements**

We enjoyed several helpful discussions with Drs. S. Bowen, C.C. Gerry, S. Hassani and R.E. Wagner. This work has been supported by the NSF. We also acknowledge support from the Research Corporation.